\documentclass[12pt,preprint]{aastex}

\def\etal{{\it et al.}}
\def\ie{{\it i.e.}}
\def\eg{{\it e.g.}}

\def\gap{\hbox{${_{\displaystyle>}\atop^{\displaystyle\sim}}$}}
\def\omegabf{\mbox{\boldmath$\omega$}}

\begin{document}

\title{Precession Interpretation of the Isolated Pulsar PSR B1828-11} 
\author{Bennett Link\altaffilmark{1}}
\affil{Montana State University, Department of Physics, Bozeman MT
59717; blink@dante.physics.montana.edu} 
\altaffiltext{1}{Also Los Alamos National Laboratory}
\and
\author{Richard I. Epstein}
\affil{Los Alamos National Laboratory, Mail Stop D436, Los Alamos, NM
87545; epstein@lanl.gov}

\begin{abstract}

Pulse timing of the isolated pulsar PSR B1882-11 shows strong Fourier
power at periods $\simeq 1000$, 500 and 250 d, correlated with changes
in the pulse profile (Stairs, Lyne \& Shemar 2000).  We study the
extent to which these data can be explained by precession of the
star's rigid crust coupled to the magnetic dipole torque. We find that
the correlated changes in the pulse duration and spin period
derivative can be explained as precession at a period of $\simeq 500$
d with a wobble angle of $\simeq 3^\circ$ if the star's dipole moment
is nearly orthogonal to its symmetry axis. The dipole torque produces
a harmonic at $\sim 250$ d. Comparison of the predicted spin dynamics
with the observed pulse durations requires the radio beam to have a
non-standard ``hour-glass'' shape. We make predictions of variations
in beam polarization and pulse profile with which to test this
interpretation. The precession interpretation of PSR B1828-11
seriously challenges the current understanding of the liquid interior
of the neutron star. In particular, if the internal liquid is in a
superfluid state, its rotational vortices cannot be significantly
pinned to the crust.
\end{abstract}


\section{Introduction}

The monitoring of pulsar spin behavior through precision timing
measurements provides a means for studying the dynamics and internal
structure of neutron stars. In this connection, detection of
precession in pulsars is of particular interest. Relativistic
(geodetic) precession has been observed in the double neutron star
binaries PSRs B1913+16 and B1534+12 (Weisberg, Romani \& Taylor 1989;
Arzoumanian 1995), and provides confirmation of General Relativistic
orbital mechanics.  Classical precession in isolated pulsars could
probe the coupling of the stellar crust to the interior, but its
detection is complicated by the presence of timing noise. Evidence for
long-period precession is seen in the Crab pulsar (Lyne, Pritchard \&
Smith 1988), the Vela pulsar (Deshpande \& McCulloch 1996) and PSR
B1642-03 (Cordes 1993; Shabanova, Lyne \& Urama 2001). The 35-d
periodicity seen in the accreting system Her X-1 (Tannanbaum \etal\
1972) has been interpreted as precession by many authors (\eg, Brecher
1972; Tr\"umper \etal\ 1986; Cadez, Galicic \& Calvani, 1997; Shakura,
Postnov \& Prokhorov 1998).

Recently, Stairs, Lyne \& Shemar (2000) discovered the most
compelling evidence to date for precession of an isolated neutron
star. The period residuals of PSR B1828-11 have an amplitude of $\sim
1$ ns and have strong Fourier power at periods of $\simeq 1000$
d, 500 d and 250 d. These Fourier components alone might suggest a
Doppler effect through gravitational interaction with planets.
However, correlations of the pulse shape and duration with
variations in the period derivative $\dot p$ strongly suggest that the
viewing angle with respect to the emission beam is varying, as one would expect
if the star were precessing (see Fig. 1). There is no obvious damping
of the observed fluctuations, though the timing behavior is not
entirely stable. See Table 1 for the spin parameters of PSR B1828-11. 

\begin{table*}[t]
\caption[]{Spin Parameters of PSR B1828-11$^a$}
      \[
         \begin{array}{ll}
            \hline
            \noalign{\smallskip}
            \hline
            \noalign{\smallskip}
\mbox{period } p & 0.405 \mbox{ s} \\
\dot{p} & 60.0 \times 10^{-15} \mbox{s s}^{-1} \\
\mbox{Spindown age }t_{\rm age}\equiv p/2\dot{p} & 0.11\mbox{ Myr} \\ \hline
            \noalign{\smallskip}
            \hline
         \end{array}
      \]
\noindent $^a$Stairs, Lyne \& Shemar 2000 \\
\noindent
\end{table*}

The purpose of this paper is to investigate if the timing behavior and
pulse profile changes of PSR B1828-11 can be quantitatively understood
in terms of stellar precession. We assume that the observed timing
residuals represent true changes in the pulsar's rotation rate. An
immediate question is which of the three observed Fourier components
corresponds to the actual precession frequency. The residuals of the
period $p$ and period derivative $\dot{p}$ show the strongest Fourier
power at $\simeq 500$ d (see Fig. 1 and Stairs, Lyne \& Shemar 2000
Fig. 3); we tentatively assume that this Fourier component represents
the precession period. The {\em wobble angle} $\theta$ between the
star's symmetry axis and its angular momentum can be estimated from
the pulse profile changes. The difference in widths between the
``wide'' and ``narrow'' beam profiles is $\simeq 2^\circ$ (see
Fig. 2), implying a wobble angle of similar magnitude for plausible
beam shapes.

This precession interpretation could be complicated by magnetospheric
processes. Precession of the star could create changes in the
magnetosphere that affect the geometry and location of the emission
region. Changes in the height of the emission region would introduce
light-travel time contributions to the pulse arrival times.  Different
regions of the magnetosphere might be viewed as the star
precesses. We do not consider these effects in this paper. 

In the next section, we give an overview of precession.  In \S 3 we
study how the electromagnetic torque, responsible for the star's spin
down (at least in part), modifies precession. In \S 4, we relate this
torque effect to the observed timing. In \S 5, we propose a model of
the the observed pulse duration variations. In \S 6 we discuss some of
our chief results. We conclude with discussion of 
observational tests of the precession interpretation.

\section{Overview}

Most treatments of pulsar precession to date have considered mainly
{\em free} precession (\eg, Goldreich 1970; Pines \& Shaham 1974;
Shaham 1977; Suto \& Iso 1985; Orford 1987; Bisnovatyi-Kogan, Mersov
\& Sheffer 1989; Glendenning 1990; Nelson, Finn \& Wasserman 1990;
Bisnovatyi-Kogan \& Kahabka 1993; Cadez, Galicic \& Calvani 1997;
Shakura, Postnov \& Prokhorov 1998; Sedrakian, Wasserman \& Cordes
1999; Jones \& Andersson 2000).  The observed variations in $p$ and
$\dot{p}$ in PSR B1828-11, however, are far too large to be explained
by free precession for a wobble angle of $\theta\sim 2^\circ$. In \S 3
we calculate the precession of a neutron star subject to a vacuum
magnetic dipole braking torque and find a solution that is consistent
with observations.

Consider first the free precession of a biaxial neutron star; the geometry
is as shown in Fig. 3.  At any instant, the symmetry axis $\hat{e}_3$,
angular momentum ${\mathbf L}$ and angular velocity $\omegabf$
span a plane.  Let the oblateness be $\epsilon\equiv(I_3-I_1)/I_1$,
where $I_3$ and $I_1$ are the largest and smallest moments of inertia
about the principal axes.  For small oblateness $\epsilon\ll 1$, the
angle between ${\mathbf L}$ and $\omegabf$ is
$\hat{\theta}\simeq\epsilon\theta\ll\theta$ (see, \eg, Jones \&
Andersson 2000). In the inertial frame, $\hat{e}_3$ and
$\omegabf$ rotate in a right-handed sense about ${\mathbf L}$
at frequency $\simeq\omega(1+\epsilon)$. Combined with this motion is
a slow {\em retrograde} motion of the body about $\hat{e}_3$ at
precession frequency $\omega_p\simeq\epsilon\omega$. 

For a precessing pulsar, the pulse emission times are defined by the
instants at which the beam center passes closest to the direction of
the observer; the observed arrival times are simply offset from the
emission times (apart from Doppler shifts from translational motion).
If the star is freely precessing, the pulse arrival times vary
sinusoidally as $\Delta t\simeq (p_0/2\pi)\theta\cot\chi\sin(\omega_p
t+\beta)$ where $p_0$ is the spin period, $\chi>\theta$ is the
inclination angle between the center of the pulsar beam and the body's
symmetry axis and $\beta$ is a phase (see Nelson, Finn \& Wasserman
1990). This variation is a purely geometrical effect, as the magnitude
of the angular velocity is unchanged by free precession. The
corresponding changes in period derivative due to this geometrical
effect are $\Delta\dot{p}\simeq
(2\pi)^{-1}(p_0\omega_p)^2\theta\cot\chi\sin(\omega_p t+\beta)$ (see
eq. \ref{dpdotmodel} below). For a precession period of 500 d and
$\chi>30^\circ$, the residuals in $\dot{p}$ are less than 0.02 s
s$^{-1}$, a factor of at least 10 smaller than observed.

The electromagnetic torque on a neutron star modifies the precession
and increases the magnitude of timing residuals for a given wobble
angle (Jones 1988; Cordes 1993; Jones \& Andersson 2000). In the body
frame of a precessing object, the angular velocity vector $\omegabf$
moves about the body's symmetry axis $\hat{e}_3$ at frequency
$\omega_p$ in a right-handed sense. For a magnetized neutron star with
a magnetic moment misaligned with respect to the symmetry axis, the
angle between the dipole moment ${\mathbf m}$ and $\omegabf$ changes
by about $\theta$ over a precession period (see eq. \ref{Thetaa}
below). The external dipole torque, which depends on this angle, thus
varies over a {\em precession} period in addition to its slow decay
over an evolutionary timescale.  For small wobble angle $\theta$,
these torque changes produce changes in the star's spin rate $\omega$
of (see eq. \ref{domdot})
\begin{equation}
{\Delta\dot{\omega}\over\omega_0} =
{\theta\over2\tau}\left[ \sin{2\chi}\cos(\omega_p t+\beta) 
+{\theta\over 2}\sin^2\chi\cos(2\omega_p t +2\beta)\right], 
\end{equation}
For simplicity we assume that the pulsed radiation is emitted along
the axis of the magnetic dipole, \ie, both the beam and the dipole
are at an angle $\chi$ with respect to the body's symmetry axis.  The
torque-induced timing variations are much larger than the geometrical
effects associated free precession (Cordes 1993; Jones 1988; see
eq. \ref{dpdotmodel} below).  In addition, and as pointed out by Jones
\& Andersson (2000), the dependence of the torque on the angle between
the star's angular velocity and its dipole moment creates a strong
harmonic of the precession frequency in the star's spin rate {\em if
the star is a nearly orthogonal rotator}. In \S 5 we apply this
picture to PSR B1828-11 and find that $\chi=89^\circ$ and a wobble
angle of $\theta\simeq 3^\circ$ can account for the 500-d Fourier component
with a harmonic at 250 d. Precession does not give the whole story,
however, as it does not explain the Fourier power at 1000 d.

As the star precesses we see different sweeps through the beam of the
pulsar, allowing us to partially map the beam morphology. As we will
show, the observed variations in pulse duration imply that the beam
has an ``hour glass'' shape, \ie, the pulse duration is longer for
sweeps farthest from the beam center.

\section{Precession Under the Dipole Torque}

Precession of a neutron star will occur if the symmetry axis of the
crust is not aligned with the angular momentum. The elastic crust has two
contributions to its oblateness: a {\em centrifugal deformation}
associated with the star's spin and a {\em Coulomb deformation}
sustained by the rigidity of the lattice (Alpar \& Pines 1985; Jones
\& Andersson 2000). The principal axis of inertia of the centrifugal
deformation follows the instantaneous spin axis of the crust, and
so does not affect the precession. In \S 6 we discuss possible origins
of Coulomb deformation. 

The period and amplitude of the crust's precession are affected by the
coupling of the crust to the liquid interior. The precession creates
time-dependent velocity differences between the crust and liquid that
vary over the star's spin period.  If the coupling time $\tau_f$
between the crust and the liquid interior is much longer than the
crust's spin period $p$, the precession will damp over $\simeq
2\pi\tau_f/p$ precession periods (Bondi \& Gold 1955; Sedrakian,
Wasserman \& Cordes 1999). All studies of damping of differential
rotation between the crust and various parts of the liquid interior
give $\tau_f\gg p$ (\eg, Alpar \& Sauls 1988; Epstein \& Baym 1992;
Jones 1992; Abney, Epstein \& Olinto 1996; Mendell 1998). To a good
approximation, therefore, the liquid interior can be treated as {\em
decoupled} from the solid (Bondi \& Gold 1955; Sedrakian, Wasserman \&
Cordes 1999); in this regime the crust precesses almost as if the
liquid interior were not there.\footnote{A precessing rigid shell
containing a non-spherical normal fluid is {\em inertially coupled} to
the fluid (Lamb 1952). The fluid, which tends to a configuration
symmetric about its rotation axis, exerts a reaction force on the
solid. The relevance of this effect to an elastic neutron star
containing superfluid neutrons and protons is unclear. Inertial
coupling would not change the precession dynamics that we calculate in
this paper, though it would change the dependence of the precession
period on the crust oblateness.}

In our calculations we treat the neutron star as consisting of a
biaxial, rigid crust afloat on a liquid core. The solid portion of the
star comprises less than $1\%$ of the star's total moment of inertia
and we regard the liquid of the inner crust as an extension of the
core liquid.  We neglect angular momentum exchange between the solid
and the liquid. For simplicity we ignore the centrifugal bulge (which
does not affect the precession) and treat the crust as perfectly
rigid.  For illustration we describe the spin dynamics of the crust
with the vacuum dipole torque of Davis \& Goldstein (1970). While
plasma effects could contribute to the spin down, a torque similar to
the vacuum torque will always be present. In a frame corotating with
the crust, the equations of motion are (Goldreich 1970; Melatos 2000)
\begin{equation}
{\mathbf I}\cdot {d\omegabf\over dt} + \omegabf\times
{\mathbf L} = {2\omega^2\over 3c^3}(\omegabf\times{\mathbf
m})\times{\mathbf m} + {1\over R c^2}(\omegabf\cdot{\mathbf
m})(\omegabf\times {\mathbf m}),
\end{equation}
where ${\mathbf I}$ is the moment of inertia tensor of the crust,
$\omegabf$ is the angular velocity, ${\mathbf L}$ is the angular
momentum, ${\mathbf m}$ is the magnetic dipole moment (centered on the
star) and $R$ is the radius of the star. The first contribution to the
torque is due to the radiation far field and is responsible for
spinning down the star. The second term represents the near-field
radiation torque. The coefficient of this term depends on how the
star's internal magnetization is treated (Melatos 2000); its precise
value does not affect our results. We will find that variations in the
spin rate associated with precession are determined predominantly by
the far-field torque (see. eq. \ref{domdot}).

We define a Cartesian coordinate system (1,2,3) with the 3 axis along
the symmetry axis $\hat{e}_3$ of the crust. Without loss of
generality, we take the magnetic dipole to lie in the 1-3 plane, 
inclined with respect to the 3 axis by an angle $\chi<\pi/2$. At $t=0$
the spin rate is $\omega_0\equiv 2\pi/p_0$. The equations of motion are
\begin{eqnarray}
I_1 {\dot\omega_1} + \omega_2\omega_3 (I_3-I_1) & = & T_1 = 
{1\over 2\tau}I_3  (\omega_3\sin\chi  -  \omega_1\cos\chi)\cos\chi \cr 
& & \qquad\qquad +{3x_0^{-1}\over 4\tau}I_3 (\omega_1\sin\chi+\omega_3\cos\chi)\omega_2\sin\chi
\label{eom1} \\
I_1 {\dot\omega_2} - \omega_1\omega_3 (I_3-I_1) & = & 
T_2 = - {1\over 2\tau}I_3 \omega_2 \cr 
& & \qquad\qquad + {3x_0^{-1}\over
4\tau}I_3 (\omega_1\sin\chi +\omega_3\cos\chi)(\omega_3\sin\chi-\omega_1\cos\chi)
\label{eom2}\\
I_3 {\dot\omega_3} & = & 
T_3 = {1\over 2\tau}I_3 \sin\chi (\omega_1\cos\chi - \omega_3\sin\chi) \cr
& & \qquad\qquad -{3x_0^{-1}\over 4\tau}I_3  (\omega_1\sin\chi +  \omega_3\cos\chi)\omega_2\sin\chi,
\label{eom3}
\end{eqnarray}
where $\tau\equiv 3c^3I_3/4m^2\omega_0^2$ is the characteristic time
over which the far-field torque spins down the star,
$m\equiv\vert{\mathbf m}\vert$ and $x_0\equiv R\omega_0/c<<1$. For the
oblate star, $I_3>I_2=I_1$. Terms on the right-hand side of
eqs. [\ref{eom1}]-[\ref{eom3}] proportional to $x_0^{-1}$ arise from
the near-field torque; the remaining terms are due to the far-field
torque. The near-field torque is stronger than the far-field torque by
a factor $\simeq x_0^{-1}$, though its direction is always orthogonal
to $\omegabf$.

Identification of the small parameters in the problem allows
perturbative treatment of the torque. Because the torque acts over
timescales that are much longer than the observed precession period,
we expect the spin solution to resemble that for free
precession. Hence, the ratio $\omega_p/\omega_0$, where $\omega_p$ is
the frequency of free precession, should be approximately
$\epsilon\equiv (I_3-I_1)/I_1$. This ratio is $\simeq 10^{-8}$ for PSR
B1828-11 if the precession period is $\simeq 500$ d. Changes in the
pulse duration (Fig. 2) through the precession period suggest a wobble
angle of $\theta\sim 2^\circ$. For these small values of $\epsilon$
and $\theta$, $\hat{\theta}\simeq\epsilon\theta\ll\theta$, so $\theta$ and
$\alpha$ are nearly equal (Jones \& Andersson 2000). In
terms of the components of the angular velocity, $\alpha$
is approximately $(\omega_1^2+\omega_2^2)^{1/2}/\omega_0\ll 1$.  We
therefore regard $\omega_1$ and $\omega_2$ as perturbations about the
secular spin down. The relative magnitudes of the small dimensionless
parameters are given by
\begin{equation}
1\gg\alpha\gg\epsilon
\gg \epsilon\alpha
>x_0^{-1} {1\over\omega_0\tau}
\gg {1\over\omega_0\tau}.
\label{ineq}
\end{equation}
To order $\epsilon\omega_1$ and $\epsilon\omega_2$, eqs. [\ref{eom1}]
and [\ref{eom2}] are
\begin{eqnarray}
\dot{\omega_1} + \epsilon \omega_2\omega_0 = 0 \\
\dot{\omega_2} - \epsilon \omega_1\omega_0 = 0.
\end{eqnarray}
For $\alpha\ll 1$, the solutions are
\begin{eqnarray}
\omega_1&=&\omega_0\alpha\cos(\omega_pt+\beta) \label{omega1}\\
\omega_2&=&\omega_0\alpha\sin(\omega_pt+\beta) \label{omega2}\\
\omega_p&=&\epsilon\,\omega_0={I_3-I_1\over I_1}\omega_0,\label{omegap}
\end{eqnarray}
where $\beta$ is a phase.  The behavior of the component of the
angular velocity orthogonal to the symmetry axis is essentially
unaltered by the torque; the precession frequency is that of free
precession and
$\alpha\simeq\sin\alpha=(\omega_1^2+\omega_2^2)^{1/2}/\omega_0=$constant. In
the body frame, both ${\mathbf L}$ and $\omegabf$ (which are always
coplanar with $\hat{e}_3$) precess slowly at frequency
$\omega_p\ll\omega_0$ about $\hat{e}_3$ and in the same sense. For the
limit $\epsilon\ll 1$ that we are considering, $\alpha$ and $\theta$
are nearly equal. 

For biaxial free precession, $\omega\equiv\vert\omegabf\vert$ is
constant. Under the dipole torque, however, $\omega$ undergoes changes
that are important for the observed timing behavior. Combining
eqs. [\ref{eom1}]-[\ref{eom3}] gives
\begin{equation}
{d\omega^2\over dt} = {2\over I_1}\left (\omegabf\cdot{\mathbf
T} - \epsilon {I_1\over I_3} \omega_3 T_3 \right ). 
\label{domega2}
\end{equation}
The $\omegabf\cdot{\mathbf T}$ term does not depend on the near-field
torque. The near-field torque changes the spin rate only through the
small final term, which we will neglect. The secular spindown in the
absence of precession is given by $\dot{\omega}\simeq
-\omega\sin^2\chi/2\tau$. Combining eq. [\ref{domega2}] with
eqs. [\ref{omega1}]-[\ref{omegap}], setting $\alpha=\theta$, retaining
terms to order $\theta^2(\omega_0\tau)^{-1}$, and subtracting the
secular spindown gives (dropping constant terms):
\begin{equation}
{\Delta\dot{\omega}\over\omega_0} =
{\theta\over2\tau}\left[ \sin{2\chi}\cos(\omega_p t+\beta) 
+{\theta\over 2}\sin^2\chi\cos(2\omega_p t +2\beta)\right]. 
\label{domdot}
\end{equation}
Hence $\dot{\omega}$ undergoes small variations of order
$\theta(\omega_0\tau)^{-1}$ and $\theta^2(\omega_0\tau)^{-1}$, driven
by changes in the far-field dipole torque. These torque variations are
due to changes in the angle between $\omegabf$ and ${\mathbf m}$ as
$\omegabf$ precesses through the body. The harmonic at twice the
precession frequency arises from quadratic dependences of $\omega_1
T_1$ and $\omega_2 T_2$ on the components of the angular velocity; the
harmonic dominates the fundamental for $\theta\tan\chi>4$. For
a small $\theta$, the dipole moment must be nearly orthogonal to
the rotation axis for the harmonic to be significant. The small
oscillations of $\dot{\omega}$ largely determine the observed pulse
timing; they give period variations $\Delta\dot
p\simeq -(p_0^2/2\pi)\Delta\dot\omega$. 
 
\section{Timing} 

The observed pulse timing at a given time in the precession cycle
depends on the star's spin rate and the orientation of the angular
velocity $\omegabf$ with respect to the observer. To relate the
solution of the previous section to the observed pulse timing, we go
to an inertial coordinate system $x,y,z$ with the angular momentum
along the $z$ axis and the observer in the $x-z$ plane with $x>0$ and
$z>0$.  The azimuthal and polar angles of the magnetic dipole
${\mathbf m}$, $\Phi$ and $\Theta$, are given in this frame by
\begin{eqnarray}
\tan\Phi & = & {m_y\over m_x} = {\left
(\cos\psi\sin\phi+\cos\theta\cos\phi
\sin\psi\right )\sin\chi-\sin\theta\cos\phi\cos\chi
\over\left (\cos\psi\cos\phi-\cos\theta\sin\phi\sin\psi\right
)\sin\chi+
\sin\theta\sin\phi\cos\chi} \label{Phi} \\
\cos\Theta & = & {m_z\over m} = \sin\theta\sin\psi\sin\chi+\cos\theta\cos\chi,
\label{Theta}
\end{eqnarray}
where $\psi$, the wobble angle $\theta$ and $\phi$ are Euler angles
following the definitions of Landau \& Lifshitz (1976) and Goldstein
(1980).
 
The pulsar beam is not necessarily in the same direction as the dipole
moment. Nevertheless, for simplicity we define a pulse as occurring
when the azimuthal angle $\Phi$ of the magnetic dipole equals the
azimuth of the observer, \ie, when $m_y=0$ and $\Phi=0$. Hence
$\dot\Phi$ is the observed pulse frequency.  For small $\theta<\chi$
and $\theta\cot\chi\ll 1$, eqs. [\ref{Phi}] and [\ref{Theta}] give to
second order in $\theta$,
\begin{equation}
{\dot\Phi}  =  \dot{\phi} + \dot{\psi} +
\theta\cot\chi\,\dot{\psi}\sin\psi
-{\theta^2\over 2}(1+2\cot^2\chi)(2\cos^2\psi-1)\dot{\psi}.
\end{equation}
The angles $\psi$ and $\phi$ are given by
\begin{eqnarray}
\psi & = & \tan^{-1} {L_1\over L_2} = \tan^{-1} {\omega_1\over \omega_2} =
{\pi\over 2}-\omega_p t \label{psi} -\beta \\
\dot{\phi} & = &
{\omega_3-\dot{\psi}\over\cos\theta}\simeq \omega + \omega_p +{1\over
2}\theta^2\omega_p,\\
\label{phi}
\end{eqnarray}
The observed beam sweep rate is (dropping constant terms) 
\begin{equation}
{\dot\Phi} = \omega - \theta\,\omega_p\cot\chi\,\cos(\omega_p t+\beta)
- \theta^2\omega_p\,(1+2\cot^2\chi) \cos^2(\omega_p t + \beta)
\label{dphidt} 
\end{equation}
Hence, the sweep rate is the spin rate plus a modulation at the free
precession frequency and its first harmonic. This modulation is a
geometrical effect due to variations in the rate at which the beam
sweeps past the observer as the symmetry axis precesses about the
angular momentum; it is the only modulation that occurs if the
precession is free.  The sweep rate also changes as $\omega$ changes
through secular spin-down and the variations $\Delta\omega$ associated
with torque variations as the star precesses. We henceforth focus on
behavior of the period derivative since, as we will show, the torque
changes evaluated in the previous section largely determine the
observe spin behavior. Differentiating eq. [\ref{dphidt}] in time and
combining with eqs. [\ref{domdot}], [\ref{psi}] and [\ref{phi}], gives
the period derivative residuals with respect to the secular spin down:
\begin{eqnarray}
\Delta\dot{p}\simeq -{p_0^2\over 2\pi}{\ddot\Phi}& \simeq &-
{p_0\over t_{\rm age}}\theta\left[ \cot\chi\cos(\omega_p t+
\beta)+{\theta\over 4}\chi\cos(2\omega_p t +2\beta)\right ] \cr
& - & {1\over
2\pi}(p_0\omega_p)^2 \theta \left[ \cot\chi\sin(\omega_p t+\beta) +
\theta(1+2\cot^2\chi)\sin(2\omega_p t+2\beta)\right].
\label{dpdotmodel}
\end{eqnarray}
Here $t_{\rm age}\equiv\tau/\sin^2\chi$ is the characteristic spin-down
age.  This equation gives the complete expression for the star's
residuals in $\dot p$. The first set of square brackets gives the
residuals due to torque variations from eq. [\ref{domdot}]. The second
set of square brackets gives the residuals due to geometrical effects
(see Nelson, Finn \& Wasserman 1990; Jones \& Andersson 2000). For PSR
B1828-11, the first-order torque effects always dominate the
geometrical effects by a factor of $\sim 250$ if the precession period
is $\simeq 500$ d. For the harmonic at frequency $2\omega_p$ to play a
role in the timing (terms proportional to $\theta^2$), $\chi$ must be
nearly $\pi/2$. The near-field torque, because it does not
significantly change $\omega$, is unimportant for the small-$\theta$
precession that appears to be taking place in PSR B1828-11.

We model the data for the 2000-day span beginning at MJD 49,000 as the
observations were most closely spaced during this period and these
data show the observed periodicities most clearly. Fitting
eq. [\ref{dpdotmodel}] to the data, neglecting the small geometrical
terms, gives the parameters $\theta$, $\omega_p$, $\beta$ and $\chi$
featured in Table 2. Comparison of the fit to the data is shown in
Fig. 1.

\begin{table*}[t]
\caption[]{Model Parameters}
      \[
         \begin{array}{ll}
            \hline
            \noalign{\smallskip}
            \hline
            \noalign{\smallskip}
\theta & 3.2^\circ \\ 
\chi & 89^\circ \\
2\pi/\omega_p &  511 \mbox{ d} \\
\xi & -0.3^\circ \\
\hline
            \noalign{\smallskip}
            \hline
         \end{array}
      \]
\noindent
\end{table*}

\section{Pulse Duration Variations} 

The duration of PSR B1828-11's pulses is seen to change with the same
periodic structure as the spin residuals, suggesting different sweeps
through the beam as the star precesses.  To study this further,
consider the behavior of the polar angle $\Theta$ of the beam with
respect to ${\mathbf L}$ (eq. \ref{Theta}). Expanding
eq. [\ref{Theta}] to first order in $\theta$, using eq. [\ref{psi}],
gives
\begin{equation}
\Theta  =  \chi - \theta \cos(\omega_p t+\beta). 
\label{Thetaa}
\end{equation}
As the star precesses, $\Theta$ varies sinusoidally about $\chi$. The
variations in the polar angle are in phase with the dominant
contribution to the torque variations (the first term in
eq. [\ref{dpdotmodel}]). 

To quantify this further, suppose the constant angle between the
observer and the angular momentum is $\gamma\equiv\xi+\chi$ (see
Fig. 4). We define the {\em sweep angle} $\Delta\Theta$ as the
difference in polar angle between the observer and the dipole at the
time of the pulse: $\Delta\Theta=\xi+\theta\cos(\omega_p t+\beta)$. As
a simple model of the variations in pulse duration, we suppose that
the pulse duration $w$ is a function of $\Delta\Theta$ only. We assume
that $w$ has an extremum (it will turn out to be a minimum) for
$\Delta\Theta=0$, corresponding to a viewing angle co-linear with the
dipole axis. We take the expansion
\begin{equation}
w = w_0 + w_2 (\Delta\Theta)^2.
\label{shape}
\end{equation}
Consistent with Stairs, Lyne \& Shemar (2000), we define a {\em shape
parameter} $S$ as 
\begin{equation}
w \equiv {\rm max}\,(w) S  + {\rm min}\,(w) (1-S), 
\end{equation}
where max $(w)$ and min $(w)$ are the maximum and minimum values of
the beam duration in the precession cycle. 
Combining eqs. [\ref{Thetaa}] and [\ref{shape}] gives 
\begin{equation}
S  = \left\{ \begin{array}{ll} 
1 - \left({\Delta\Theta\over\vert\xi\vert+\theta}\right)^2
& \mbox{$w_2>0$} \cr
\left({\Delta\Theta\over\vert\xi\vert+\theta}\right)^2
& \mbox{$w_2<0$}.
\end{array}
\right.
\end{equation}
This shape parameter depends on $\xi$, $\theta$, $\omega_p$, 
$\beta$, and the sign of $w_2$ but not its magnitude. 

With $\theta$, $\omega_p$ and $\beta$ given by the fit to the timing
data, the shape parameter is determined entirely by $\xi$ and the sign
of $w_2$. Interestingly, we find that $w_2$ must be positive, \ie, the
beam duration is larger for beam sweeps farthest from the dipole axis.
Such a beam pattern is not standard, but is {\em required} by the
precession interpretation of PSR B1828-11.  The best-fit value of
$\xi$ is $-0.3^\circ$, as given in Table 2.  Comparison of the fit to
the data is shown in Fig. 1.

\section{Discussion}

We find an acceptable fit to the data with $\chi=89^\circ$ between the
magnetic dipole and the star's symmetry axis, a precession period of
511 d and a wobble angle $\theta=3.2^\circ$. The predicted shape changes
are in qualitative agreement with the observations if the angle
between the observer and the angular momentum is $88.7^\circ$ and
the radio beam pattern has the ``hourglass'' shape sketched in
Fig. 5. The beam shape that we have inferred differs form the usual
picture of a roughly circular pulsar beam, but might be explainable in
terms of a patchy beam of the type found by Han \& Manchester (2000)
in their statistical study of pulsar beam morphology. If the emission
is beamed both parallel and anti-parallel to the dipole axis, an inner
pulse would be expected for the nearly-orthogonal inclination that we
have inferred. None is seen, so the emission pattern must be more
complex.

Our model does not account for 1000-d Fourier component seen in the
data. However, significant aperiodic variations in the timing seen
over timescales of years indicate that more than just precession is at
work in PSR B1828-11. Slow changes in the emission region could be a
contributing factor. Perturbations to the star's figure brought about
by relaxation of the star's structure as it spins down could also play
a role. Such starquakes would excite precession by perturbing the
star's angular velocity with respect to the symmetry axis (Link,
Franco \& Epstein 1998). The perturbations would introduce irregular
contributions to the magnitude of the wobble angle and the
precessional phase (Jones 1988).

Andersson \& Jones (2000) have suggested that the 1000-d Fourier
component represents the precession period, and that near
orthogonality of the dipole moment with respect to the symmetry axis
gives a harmonic at 500 d in the spin residuals. This scenario is
unlikely for several reasons: 1) near perfect orthogonality ($\chi\gap
89.9^\circ$) is required to keep the first-order torque term in
eq. [\ref{dpdotmodel}] from giving too large a contribution to
$\dot{p}$, 2) the strong Fourier component at 250 d is not accounted
for, and, 3) the shape parameter, which would now be periodic at 1000
and 500 d, would bear little resemblance to the data.

The stellar oblateness inferred from the precession period of 511 d is
$\epsilon=(I_3-I_1)/I_1=9\times 10^{-9}$. For comparison, we estimate
the maximum Coulomb oblateness. The maximum mass $M_m$ of a
``mountain'' that the rigid neutron star crust can support is given
roughly by $M_m g \sim B\theta_c/R$ where $g$ is the gravitational
acceleration at the stellar surface, $B$ is the bulk modulus of the
crust ($\simeq 10^{48}$ erg) and $\theta_c$ is the critical strain
angle at which the crust fractures. The maximum oblateness sustainable
by crust rigidity is thus (see also Ushomirsky, Cutler \& Bildsten 2000)
\begin{equation}
\epsilon\equiv {I_3-I_1\over I_1} \simeq {M_m\over M_c} \sim
10^{-5} {\theta_c\over 10^{-2}},
\end{equation}
where $M_c$ is the mass of the crust. A mountain could form as the
result of a starquake that perturbs the star's principal axis
of inertia away from its angular momentum axis (Link, Franco \&
Epstein 1998). However, even if the crust is relaxed (\ie, unstrained)
before it is set into precession, rigidity prevents a portion of the
star's bulge from following the instantaneous rotation axis. The
effective oblateness due to rigidity is in this case $\epsilon_r\simeq
10^{-5}\epsilon_\Omega$, where $\epsilon_\Omega$ is the centrifugal
oblateness (Alpar \& Pines 1985; see also Munk \& MacDonald 1960). For
PSR B1828-11's spin period $\epsilon_r \simeq 2\times 10^{-11} (p/0.4
\mbox{ s})^{-2}$, giving a precession period of 700 yr. We conclude
that PSR B1828-11's crust is strained to the extent that the Coulomb
oblateness is $\simeq 100$ times larger than $\epsilon_r$. A
significantly strained crust can precess at a much higher frequency
than the unstrained crust considered in some previous studies (\eg,
Goldreich 1970 and Melatos 2000).

The precession of PSR B1828-11 seriously challenges the notion that
superfluid vortices of the crust pin to nuclei (\eg, Anderson \& Itoh
1975). Pinning would drastically alter the way in which the star
precesses. Shaham (1977) showed that very effective pinning changes
the precession period to $(I_1/I_p)p$, where $I_p$ is the
portion of the star's fluid moment of inertia that is pinned and
$I_1$ is the moment of inertia of the part of the star
that is precessing. Sedrakian, Wasserman \& Cordes (1999) have shown
that Shaham's conclusion is essentially unaltered if the pinning is
imperfect. Pulsar glitches, which might arise from variable coupling
between the superfluid and the crust, indicate that $I_p/I_{\rm
star}>1.4$\% in stars that frequently glitch (Link, Epstein \&
Lattimer 1999); $I_{\rm star}$ is the total moment of inertia of the
star. This degree of pinning in PSR B1828-11 would give a precession
period of $\ll 40$ s, far shorter than the observed of 500 d
precession period.  It may be that vortex pinning is inhibited in
stars that undergo precession with amplitudes as large as those in PSR
B1828-11.

\section{Observational Tests}

Our proposed model for the fluctuations in the timing residuals of PSR
B1828-11 implies that the pulsed emission should have several specific
attributes that may be observationally tested. Fig. 5 illustrates
these features that are characteristic of our model. During the 511
day precession cycle the observer sees both the broad upper and lower
parts of the beam. One important test of our model would be
evidence of these two different parts of the emission region.

We modeled the observed pulse durations with the parabolic form of
eq. (\ref{shape}), giving the overall beam shape illustrated by the dashed
lines in Fig. 5. The actual shape of the emission must be more complex,
however, as we suggest with the irregular brightness contours also
shown in the figure. Additionally, the linear polarization of the
emission region is unlikely to be the same in the upper and lower
portions of the emission region, as indicated by the double- headed arrows
in the figure.

To indicate the possible signatures one may observe, we have marked
certain viewing angles (A-C) in Fig. 5 and the corresponding observing
times in Fig. 1. The dotted horizontal line in Fig. 5 is the shape
equator (perpendicular to the largest principal moment). The
horizontal dotted lines A and C give the range of the viewing angles
seen from Earth.  Line B represents the viewing angle at which the
pulse profile is the narrowest. The pulse profile is wider at viewing
angle A and widest at viewing angle C.

While the shape parameter is the same at phases above and below phase
B, the precise pulse profiles and radio polarization should be
noticeably different at these phases. The polarization angle will vary
across the emission region so that the polarization direction will
change continuously from phase A to C and then reverse going from C to
A. Detection of these changes in pulse shape and polarization would
indicate that the radio beam has the hour-glass shape we have
inferred, and would support the interpretation that PSR B1828-11 is a
precessing neutron star.

\acknowledgements We thank A. G. Lyne, D. I. Jones, R. Bandiera,
A. Melatos, J. Cordes, D. Chernoff and the other participants of the
ITP workshop on {\sl Spin and Magnetism in Young Neutron Stars} for
helpful discussions. We are grateful to I. Stairs and A. G. Lyne for
providing us with the timing data and beam templates for PSR B1828-11.
The work was performed under the auspices of the U.S. Department of
Energy, and was supported by NASA ATP grant NAG 53688, by IGPP at LANL
and in part by the National Science Foundation under Grant
No. PHY94-07194.

\newpage

\begin{figure}[t]
\plotone{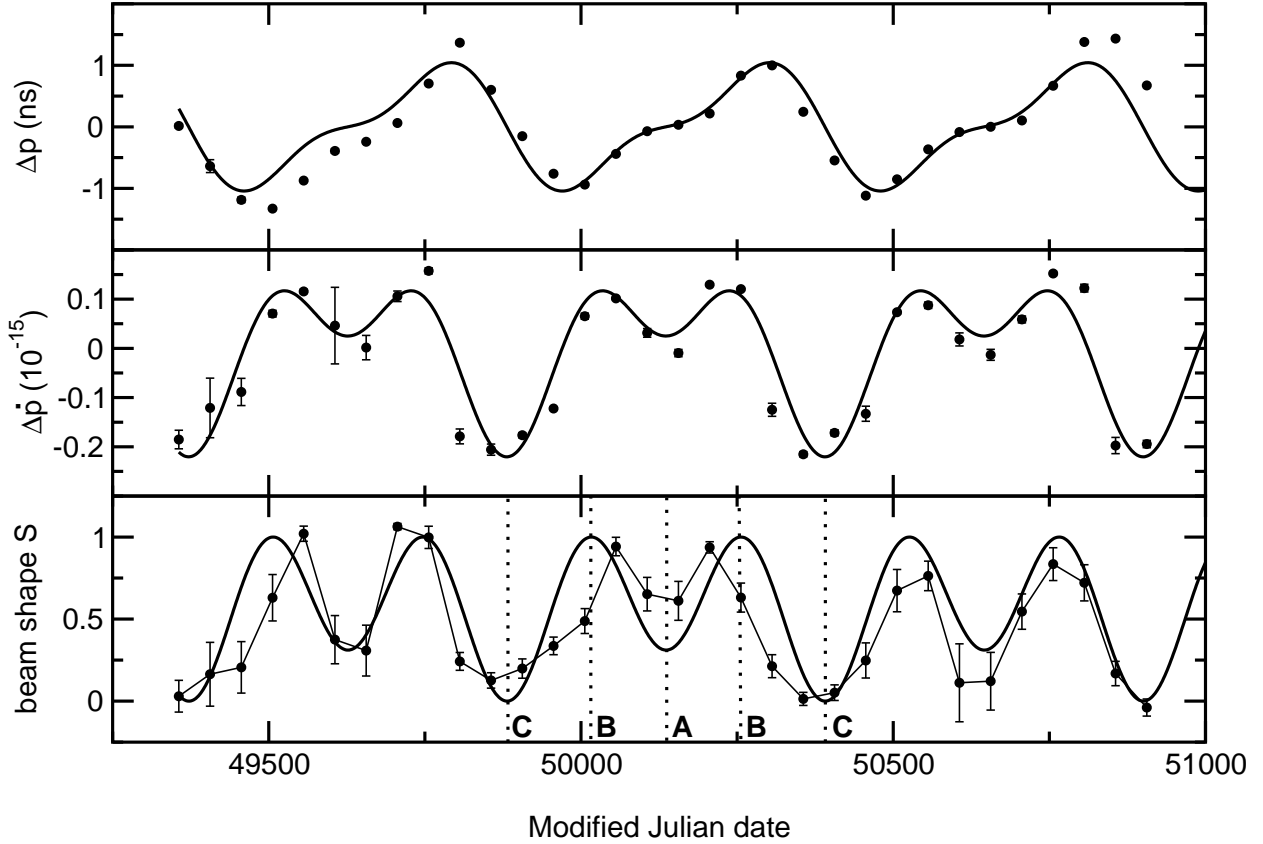}
\caption{Timing and beam shape data for PSR B1828-11 (from Stairs,
Lyne \& Shemar 2000). Only the most-densely sampled portion of the
13-year data span is shown. The top panel gives the period residuals
with respect to the star's secular spin down. The middle panel is the
time derivative. The bottom panel shows the shape parameter of Stairs,
Lyne \& Shemar (2000); it is $S=A_N/(A_N+A_W)$, where $A_N$ and $A_W$
are the fitted heights of the narrower and wider profiles, so that
$S\simeq 0$ for wide pulses and $S\simeq 1$ for narrow ones. The data
points give average values of $S$, obtained by averaging $S$ over
multiple bins. The solid curves for $\Delta p$ and $\Delta\dot{p}$ are
the fit described in the text. The jagged line in the lower panel is a
guide to the eye.}
\end{figure}

\begin{figure}[t]
\plotone{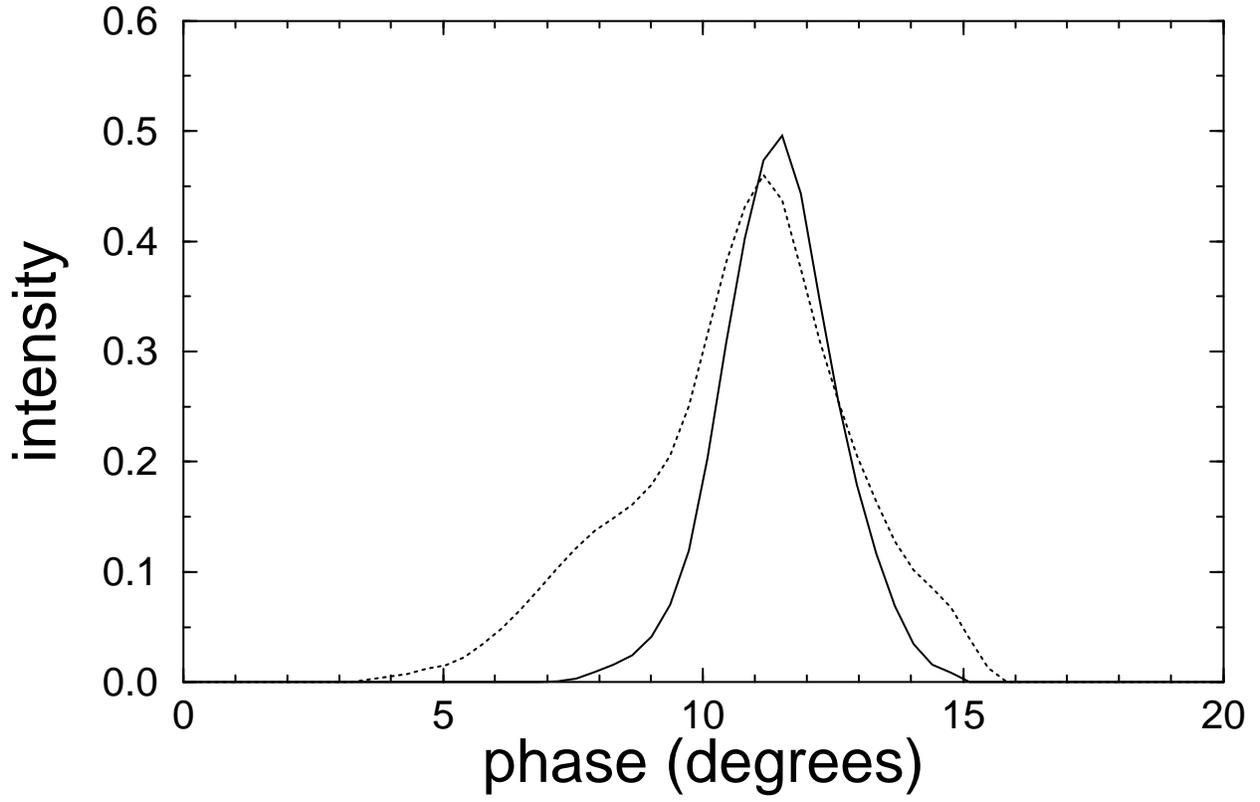}
\caption{Templates for the wide and narrow beam profiles of PSR
B1828-11 (courtesy I. Stairs). }
\end{figure}

\begin{figure}[c]
\epsscale{0.5}
\plotone{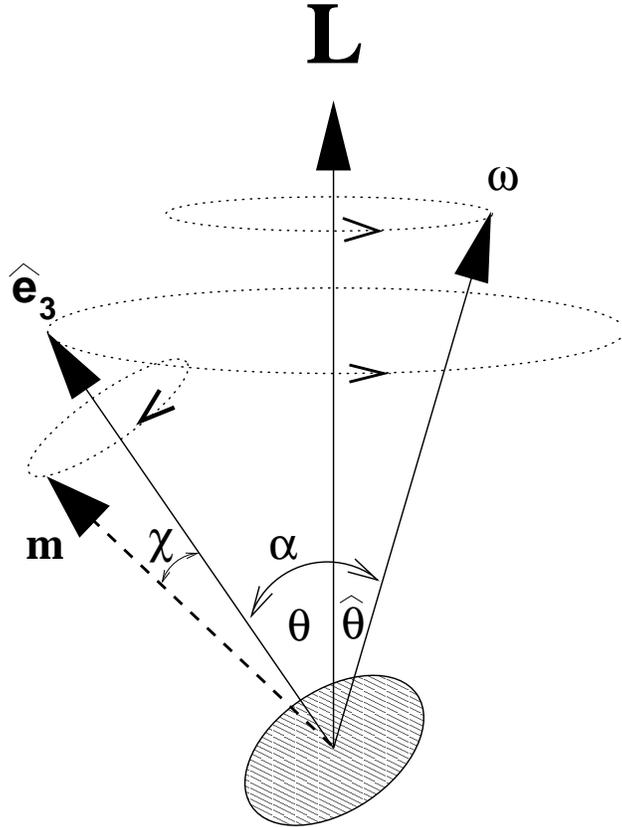}
\caption{Geometry of free precession of a biaxial neutron star in the
inertial frame. The body's symmetry axis is denoted by $\hat{e}_3$,
the angular momentum by ${\mathbf L}$ and the angular velocity by
$\omegabf$; the three vectors always span a
plane as shown. The angles $\theta$, $\hat{\theta}$ and $\alpha$ are
constant, with $\hat{\theta}\simeq\epsilon\theta\ll\theta$ for small
oblateness $\epsilon$. The vectors $\hat{e}_3$ and
$\omegabf$ rotate about ${\mathbf L}$ at nearly the spin frequency
$\omega$.  A dipole moment ${\mathbf m}$ fixed in the body, and taking an
angle $\chi$ with respect to $\hat{e}_3$, rotates in a retrograde
sense about $\hat{e}_3$ at frequency$\simeq\epsilon\omega$.}
\end{figure}

\begin{figure}[t]
\plotone{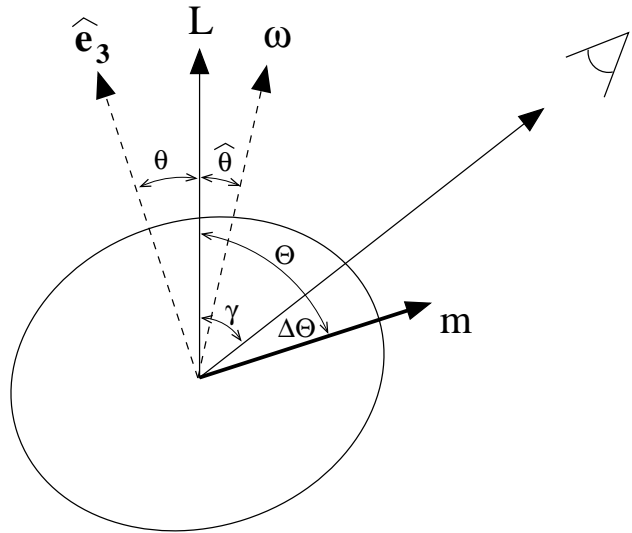}
\caption{Observing geometry. At the instant the dipole moment
${\mathbf m}$ is in the plane containing the angular momentum
${\mathbf L}$ and the observer, the angles are defined as shown.}
\end{figure}

\newpage

\begin{figure}[t]
\plotone{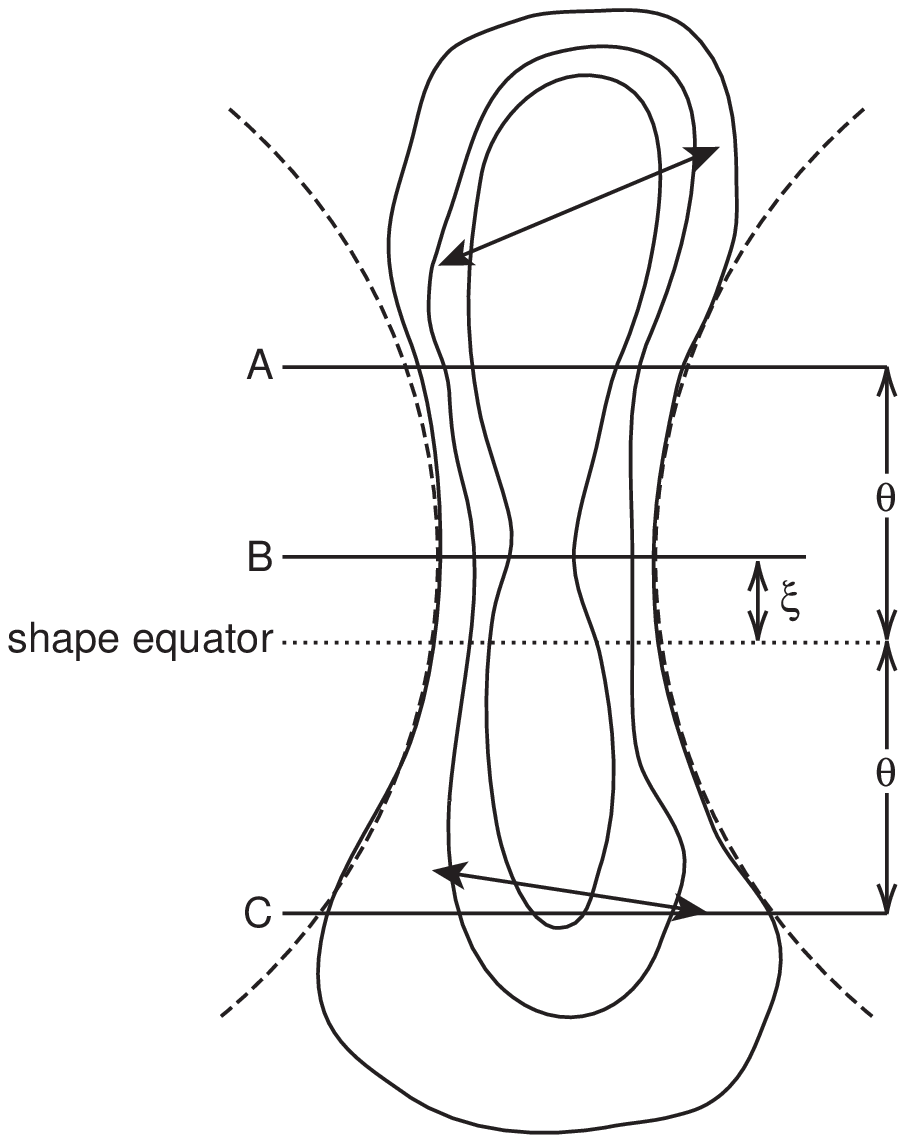}
\caption{The beam pattern for the radio emission from PSR
B1828-11.  The dashed parabolas represent the analytic fit to the beam
width, and the contour curves are one possible realization of the
actual beam shape. As the star precesses, observer's line of sight
varies by $\pm \theta$ about the star's shape equator, the dotted
horizontal line. The solid lines A and C give the range of the viewing
angles.  Line B represents the viewing angle at which the beam is the
narrowest. The double-headed arrows represent plausible polarization
directions.}
\end{figure}


\newpage

\begin{thebibliography}{}

\bibitem[a]{b}
Abney, M., Epstein, R. I. \& Olinto, A. V. 1996, \apj, 466, L91. 

\bibitem[c]{d}
Alpar, M. A. \& Pines, D. 1985, Nature, 314, 334. 

\bibitem[d]{f}
Alpar, M. A. \& Sauls, J. A. 1988, \apj, 327, 723.

\bibitem[g]{h}
Anderson, P. W. \& Itoh, N. 1975, Nature, 256, 25.

\bibitem[i]{j}
Arzoumanian, Z. 1995, Ph. D Thesis, Princeton University. 

\bibitem[k]{l}
Bisnovatyi-Kogan, G. S. \& Kahabka, P. 1993, A\&A, 267, L43. 

\bibitem[m]{n}
Bisnovatyi-Kogan, G. S., Mersov, G. A. \& Sheffer, E. K. 1989, A\&A,
221, L7.

\bibitem[o]{p}
Bondi, H. \& Gold, T. 1955, \mnras, 115, 41. 

\bibitem[q]{r}
Brecher, K. 1972, Nature, 239, 325. 

\bibitem[s]{t}
Cadez, A., Galicic, M. \& Calvani, M. 1997; A\&A, 324, 1005

\bibitem[u]{v}
Cordes, J. 1993, in {\sl Planets Around Pulsars}, ASP Conference
Series, Vol. 36, pp. 43-60 (Ed: Phillips, Thorsett \& Kulkarni). 

\bibitem[w]{x}
Davis, L. \& Goldstein, M. 1970, \apj, 159, L81.

\bibitem[y]{z}
Desphande, A. A. \& McCulloch, P. M. 1996, in ASP Conference
Series, Vol. 105, p. 101 (Ed: Johnston, Walker \& Bailes). 

\bibitem[aa]{bb}
Epstein, R. I. \& Baym, G. 1992, \apj, 387, 276. 

\bibitem[cc]{dd}
Glendenning, N. K. 1990, \apj, 359, 186.

\bibitem[ee]{ff}
Goldreich, P. 1970, \apj, 160, L11.

\bibitem[gg]{hh}
Goldstein, H. 1980, {\sl Classical Mechanics}, p. 143,
(Addison-Wesley). 

\bibitem[ii]{jj}
Han, J. L. \& Manchester, R. N. 2000, astro-ph/0010538.

\bibitem[kk]{ll}
Jones, P. B. 1988, \mnras, 235, 545. 

\bibitem[mm]{nn}
Jones, P. B. 1992, \mnras, 257, 501. 

\bibitem[oo]{pp}
Jones, D. I. \& Andersson, N. 2000, astro-ph/0011063. 

\bibitem[qq]{rr}
Lamb, H. 1952, {\sl Hydrodynamics} (Cambridge University Press). 

\bibitem[ss]{tt}
Landau, L. \& Lifshitz, E. M. 1976, {\sl Mechanics}, pp. 110-111,
(Pergamon). 

\bibitem[uu]{vv}
Link, B., Epstein, R. I. \& Lattimer, J. M. 1999, \prl, 83, 3362. 

\bibitem[ww]{xx}
Link, B., Franco, L. M. \& Epstein, R. I. 1998, \apj, 508, 838. 

\bibitem[yy]{zz}
Lyne, A. G., Pritchard, R. S. \& Smith, F. G. 1988, \mnras, 233, 667. 

\bibitem[aaa]{bbb}
Melatos, A. 2000, \mnras, 313, 217.

\bibitem[ccc]{ddd}
Mendell, G. 1998, \mnras, 296, 903. 

\bibitem[eee]{fff}
Munk, W. \& MacDonald, G. J. F. 1960, {\sl The Rotation of the Earth}
(New York: Cambridge University Press). 

\bibitem[ggg]{hhh}
Nelson, R. W., Finn, L. S. \& Wasserman, I. 1990, \apj, 348, 226.

\bibitem[iii]{jjj}
Orford, K. J. 1987, Ap\&SS, 129, 181. 

\bibitem[kkk]{lll}
Pines, D. \& Shaham, J. 1974, {\sl Comments on Modern Physics, Part
C}, 6, 37. 

\bibitem[mmm]{nnn}
Sedrakian, A., Wasserman, I. \& Cordes, J. M. 1999, \apj, 524, 341. 

\bibitem[ooo]{ppp}
Shabanova, T. V., Lyne, A. G. \& Urama, J. O. 2001, astro-ph/0101282.

\bibitem[qqq]{rrr}
Shaham, J. 1977, \apj, 214, 251. 

\bibitem[sss]{ttt}
Shakura, N. I, Postnov, K. A. \& Prokhorov, M. E. 1998, A\&A, 331, L37. 

\bibitem[uuu]{vvv}
Stairs, I. H., Lyne, A. G. \& Shemar, S. L. 2000, Nature, 406, 484.

\bibitem[www]{xxx}
Suto, Y. \& Iso, K.-I. 1985, Ap\&SS, 115, 243. 

\bibitem[yyy]{zzz}
Tannanbaum, H., Gursky, H., Kellog, E. M, Levinson, R., Schreier, E.,
Giacconi, R. 1972, \apj, 174, L143. 

\bibitem[1]{2}
Tr\"umper, J., Kahabka, P, \"Ogelman, H., Pietsch, W., Voges, W. 1986,
\apj, 300, L63. 

\bibitem[3]{4}
Ushomirsky, G., Cutler, C. \& Bildsten, L. 2000, astro-ph/0001129.

\bibitem[5]{6}
Weisberg, J. M., Romani, R. W. \& Taylor, J. H. 1989, \apj, 347,
1030. 

\end{thebibliography}
\end{document}